\documentclass[12pt]{article}
 \usepackage[cp1251]{inputenc}
 \usepackage[english, russian]{babel}
 \usepackage{amsmath, latexsym, amssymb, bm, array, graphics, eucal,
 amsfonts}
 \makeatletter
 \newcommand{\diag}{\rm \diag\, }
 
 \newcommand{\const}{\mathop{\rm const\, }}

\newcommand{\mc}[1]{\mathcal{#1}}
 \makeatother
 
\usepackage[dvips]{graphicx}

\begin{document}
 \renewcommand{\abstractname}{\, Abstract}
 \thispagestyle{empty}
 \large
 \renewcommand{\refname}{\begin{center}
 REFERENCES\end{center}}
\newcommand{\E}{\mc{E}}
 \makeatother

\begin{center}
\bf The kinetic Holway---Shakhov equation
\end{center}

\begin{center}
  \bf A. V. Latyshev\footnote{$avlatyshev@mail.ru$} and
  A. A. Yushkanov\footnote{$yushkanov@inbox.ru$}
\end{center}\medskip

\begin{center}
{\it Faculty of Physics and Mathematics,\\ Moscow State Regional
University, 105005,\\ Moscow, Radio str., 10--A}
\end{center}\medskip

\begin{abstract}
The new generalized kinetic equation is offered.
This equation represents a hybrid of Shakhov equation  and
ellipsoidal statistical equation of Holway.
Equation constants are expressed through such physically
significant quantities, as
viscosity of gas, its heat conductivity and self-diffusion coefficient.
Then these quantities are expressed through integral brackets.

{\bf Key words:} Shakhov's equation, ellipsoidal statistical
Holway's equation, rarefied gas, constans of equation, Ferziger's number,
Prandtl number, gas macro\-pa\-rameters.

PACS numbers: 05.20. Dd Kinetic theory, 51.10.+y Kinetic and
transport theory  of gases, 47.45.Ab Kinetic theory of gases
\end{abstract}

\begin{center}
\bf  Introduction
\end{center}

In the present work the new generalized kinetic equation is
entered. This equation represents a hybrid of Shakhov's equation
\cite{Shakhov} and ellipsoi\-dal statistical Holway's equation
\cite{Holway1}, \cite{Holway2}.

Such equation was entered already in works
\cite{Lat90} -- \cite{Lat92}. However, in these works constants
of equations have not been expressed through physically significant
macroparameters, and were considered as numerical parame\-teres
of Shakhov---Holway equa\-tions, connected only with Prandtl
number.

In work \cite{Lat90} for  Holway---Shakhov
equation the classical problem of the kinetic theory
(a Kramers' problem about isothermal sliding) has been solved.
Its decision was presented in quadratures.

In work \cite{Lat91a}
solution of boundary value Rie\-mann---Hilbert problem  from
theory of functions of complex variable  was presented.
To this problem the solution of  known Smoluchovsky problem
about temperature jump is reduced.

In work \cite{Lat91b} specification of the solution of Kramers'
problem from \cite {Lat90} is given and the solution of
thermal sliding problem also is presented.

In work \cite{Lat92} the solution of  Smoluchovsky'  problem has been given
for ellipsoidal statistical equation. In this work
the numerical parameter mattered, leading to correct Prandtl number
for model of the rarefied gas consisting from molecules---solid spheres.

Equation constants are expressed through such physically
significant quantities, as
viscosity of gas, its heat conductivity and self-diffusion coefficient.

\begin{center}
  {\bf 1. Kinetic Holway---Shakhov equation}
\end{center}

Let's consider the hybrid of linear Shakhov equation (see, for example,
\cite{Shakhov}) and ellipsoidal statistical Holway equation
(see \cite {Holway2})

$$
\mathbf{v}\dfrac{\partial h}{\partial \mathbf{r}}=-
\nu h(\mathbf{r},\mathbf{v})+\nu
\dfrac{\delta n(\mathbf{r})}{n_0}+
2\nu \Big(\dfrac{m}{2kT_0}\Big) \mathbf{v}\mathbf{u}(\mathbf{r})+
$$

$$
+\nu \Big(\dfrac{mv^2}{2kT_0}-\dfrac{3}{2}\Big)
\dfrac{\delta T(\mathbf{r})}{T_0}+
\nu \gamma \sqrt{\dfrac{m}{2kT_0}}\mathbf{v}\Big(\dfrac{mv^2}{2kT_0}-
\dfrac{5}{2}\Big)\mathbf{Q}(\mathbf{r})+
$$

$$+
\nu \omega\Big(\dfrac{m}{2kT_0}\Big)
\sum\limits_{i,j=1}^{3}
\Big(v_iv_j-\dfrac{\delta_{ij}}{3}v^2\Big)P_{ij}(\mathbf{r}).
\eqno{(1.1)}
$$ \smallskip

Parameters $\nu, \gamma, \omega$ are number parame\-ters of equation,
thus the para\-meter $\nu $ makes sense as effective collision frequency
of gas molecules.

At $ \omega=0$ the equation (1.1) transforms to Shakhov's equation, and
at $ \gamma=0$ transforms to ellipsoidal statistical Holway's equation.

Function $h({\bf r}, {\bf v}) $, named more low
the distribution function, is relative perturbation
to absolute Maxwellian $f_0(v)$,
$$
f_0(v)=n_0\Big(\dfrac{m}{2kT_0}\Big)^{3/2}\exp\Big(-\dfrac{mv^2}
{2kT_0}\Big),
$$
and quantities $n_0, T_0$ are concentration (number density)
and  gas tempe\-ra\-ture in some point $ {\bf r_0}$ of gas volume.

At the solution of concrete half-space problems this
point gets out in some point of a surface, for example,
in the origin of coordinates.

Besides, in (1.1) following designations are accepted,
$ \delta n (\mathbf {r}) =n (\mathbf {r})-n_0$ is the deviation local
concentration of gas from values $n_0$,

$$
\dfrac{\delta n(\mathbf{r})}{n_0}=
\dfrac{\beta^{3/2}}{\pi^{3/2}}\int \exp(-\beta v'^2)
h(\mathbf{r},\mathbf{v'})d^3v',
\eqno{(1.2)}
$$ \smallskip

where  $\beta=m/(2kT_0)$, further
$\mathbf{u}(\mathbf{r})$ is the mass gas velocity,
$\mathbf{U}(\mathbf{r})$ is the dimensionless mass gas velocity,
$\mathbf{U}(\mathbf{r})=\sqrt{\beta}\mathbf{u}(\mathbf{r})$,

$$
\mathbf{u}(\mathbf{r})=
\dfrac{\beta^{3/2}}{\pi^{3/2}}\int \exp(-\beta v'^2)\mathbf{v'}
h(\mathbf{r},\mathbf{v'})d^3v',
\eqno{(1.3)}
$$ \smallskip

$\delta T(\mathbf{r})=T(\mathbf{r})-T_0$
is the local deviation of gas temperature from
some value $T_0$,

$$
\dfrac{\delta T(\mathbf{r})}{T_0}=\dfrac{2\beta^{3/2}}{3\pi^{3/2}}
\int \exp(-\beta v'^2)
\Big(\beta v'^2-\dfrac{3}{2}\Big)h(\mathbf{r},\mathbf{v'})d^3v',
\eqno{(1.4)}
$$  \smallskip
$\mathbf{Q}(\mathbf{r})$ is the dimensionless vector of heat
stream,

$$
\mathbf{Q}(\mathbf{r})=\dfrac{\beta^{2}}{\pi^{3/2}}
\int \exp(-\beta v'^2)\mathbf{v'}
\Big(\dfrac{m v^2}{2}-\dfrac{5kT_0}{2}\Big)h(\mathbf{r},\mathbf{v'})d^3v',
\eqno{(1.5)}
$$ \smallskip
$P_{ij}(\mathbf{r})$ are these dimensionless components
 of viscous stress tensor,

$$
P_{ij}(\mathbf{r})=\dfrac{\beta^{5/2}}{\pi^{3/2}}
\int \exp(-\beta v'^2)\Big(v_i'v_j'-\dfrac{\delta_{ij}}{3}v'^2\Big)
h(\mathbf{r},\mathbf{v'})d^3v'.
\eqno{(1.6)}
$$ \smallskip

Let's transform the equation (1.1), having entered dimensionless
velo\-city of gas molecules ${\bf C}=\sqrt{\beta}{\bf v}$. We
have

$$
\mathbf{C}\dfrac{\partial h}{\partial \mathbf{r}}=
\nu\sqrt{\beta}\Big[-h(\mathbf{r},\mathbf{C})
+\dfrac{\delta n(\mathbf{r})}{n_0}+
2\mathbf{C}\mathbf{U}(\mathbf{r})+
$$

$$
+\Big(C^2-\dfrac{3}{2}\Big)
\dfrac{\delta T(\mathbf{r})}{T_0}+
\gamma \mathbf{C}\Big(C^2-\dfrac{5}{2}\Big)\mathbf{Q}(\mathbf{r})+
$$

$$
\omega \sum\limits_{i,j=1}^{3}\Big(C_iC_j-\dfrac{\delta_{ij}}{3}C^2\Big)
P_{ij}(\mathbf{r})\Big].
\eqno{(1.7)}
$$

Here $\mathbf{r}$ is the dimension coordinate.

The right part of the equation (1.7) is the linear integral of collisions.
In the equation (1.7)

$$
\dfrac{\delta n(\mathbf{r})}{n_0}=\dfrac{1}{\pi^{3/2}}\int \exp(-C'^2)
h(\mathbf{r},\mathbf{C'})d^3C',
\eqno{(1.2')}
$$

$$
\mathbf{U}(\mathbf{r})=\sqrt{\beta}\mathbf{u}(\mathbf{r})=
\dfrac{1}{\pi^{3/2}}\int \exp(-C'^2)\mathbf{C'}
h(\mathbf{r},\mathbf{C'})d^3C',
\eqno{(1.3')}
$$

$$
\dfrac{\delta T(\mathbf{r})}{T_0}=\dfrac{2}{3\pi^{3/2}}\int \exp(-C'^2)
\Big(C'^2-\dfrac{3}{2}\Big)h(\mathbf{r},\mathbf{C'})d^3C',
\eqno{(1.4')}
$$

$$
\mathbf{Q}(\mathbf{r})=\dfrac{1}{\pi^{3/2}}\int \exp(-C'^2)\mathbf{C'}
\Big(C^2-\dfrac{5}{2}\Big)h(\mathbf{r},\mathbf{C'})d^3C',
\eqno{(1.5')}
$$

$$
P_{ij}(\mathbf{r})=\dfrac{1}{\pi^{3/2}}\int \exp(-C'^2)
\Big(C_i'C_j'-\dfrac{\delta_{ij}}{3}C'^2\Big)
h(\mathbf{r},\mathbf{C'})d^3C',
\eqno{(1.6')}
$$

Let's notice, that a dimensionless heat stream
is connected with a dimensional heat stream

$$
\mathbf{q}(\mathbf{r})=n_0\dfrac{\beta^{3/2}}{\pi^{3/2}}\int
\exp(-\beta v^2)\mathbf{v}\Big(\dfrac{mv^2}{2}-\dfrac{5kT}{2}\Big)
h(\mathbf{r}, \mathbf{v})d^3v
$$  \smallskip
by the following equality

$$
\mathbf{q}(\mathbf{r})=\dfrac{n_0kT_0}{\sqrt{\beta}}\mathbf{Q}
(\mathbf{r}).
$$ \smallskip

And components of dimensionless of viscous stress tensor are
connec\-ted with dimension components

$$
p_{ij}(\mathbf{r})=\dfrac{n_0m}{\beta}\dfrac{\beta^{5/2}}{\pi^{3/2}}
\int \exp(-\beta v'^2)
\Big(v_i'v_j'-\dfrac{\delta_{ij}}{3}v'^2\Big)
h(\mathbf{r},\mathbf{v'})d^3v',
$$ \medskip
by the following equality
$$
p_{ij}(\mathbf{r})=\dfrac{mn_0}{\beta}P_{ij}(\mathbf{r}).
$$

Let's notice, that gas macroparameters through dimensional
velo\-city are expressed by equalities (1.2) -- (1.6), and through
dimensi\-on\-less velocity are expressed by
equalities $(1.2') - (1.6')$.

\begin{center}
\bf  2. Physical sense of parameters of the equation
\end{center}

At first we will express parameter $ \nu $ through coefficient
of self-diffusion.

Let gas consists of two identical kinds of molecules with
concentrations $n_1$ and $n_2$. For simplicity we will accept,
that $n_1\ll n_2$.
And let concen\-t\-ra\-tion of the first components varies along an
axis $x$. In considered conditions in linear approach
the second a component will be in equi\-lib\-rium with zero velocity.

For the first component the kinetic equation (1.7) will become
$$
C_x\dfrac{1}{n_1(x)}\dfrac{d n_1(x)}{d x}=-\nu\sqrt{\beta} h_1(x,{\bf C}).
$$

The solution of this equation is obvious
$$
h_1(x,{\bf C})=-C_x\dfrac{1}{\nu\sqrt{\beta} n_1(x)}\dfrac{d n_1(x)}{d x}.
$$

The diffusion stream $J_x$ equals

$$
J_x=\dfrac{n_1}{\sqrt{\beta}}\dfrac{1}{\pi^{3/2}}\int
e^{-C^2}C_xh_1(x,{\bf C})d^3C=$$$$=
-\dfrac{1}{2\nu\beta}\dfrac{d n_1(x)}{d x}=-D\dfrac{d n_1(x)}{d x},
$$
where $D$ is the self-diffusion coefficient
$$
D=\dfrac{1}{2\nu\beta}=\dfrac{kT}{m\nu}.
$$

Hence, the parameter $ \nu $ is defined by the equation
$$
\nu=\dfrac{kT}{m D}.
\eqno{(2.1)}
$$

Let's pass to  finding  the parameter $ \omega $.
Let's consider viscosity of one-atomic gas.
Let there is a gradient of $y$ - components
of mass gas velocity $u_y(x) $ along an axis $x$.
Then from the equation (1.7) we receive
$$
2\sqrt{\beta}\,C_xC_y\dfrac{d u_y(x)}{d x}=
\nu\sqrt{\beta}(-h(x,{\bf C})+2\omega C_xC_yP_{xy}).
$$

From here we find
$$
h(x,{\bf C})=2(-\dfrac{1}{\nu}C_xC_y\dfrac{d u_y(x)}{d x}+\omega
C_xC_yP_{xy}),
$$
where
$$
P_{xy}=\dfrac{1}{\pi^{3/2}}\int \exp(-C^2)C_xC_yh(x,{\bf C})d^3C,
$$
or, in explicit form,
$$
P_{xy}=\dfrac{2}{\pi^{3/2}}
\int \exp(-C^2)(-\dfrac{1}{\nu}C_xC_y\dfrac{du_y(x)}{d x}+
\omega C_xC_yP_{xy})C_xC_yd^3C.
$$

From last equation we find
$$
P_{xy}=-\dfrac{1}{2-\omega}\dfrac{1}{\nu}\dfrac{du_y(x)}{d x}.
$$
Therefore, the function $h$ is constructed
$$
h(x,{\bf C})=-2\Big(1+\dfrac{\omega}{2-\omega}\Big)
C_xC_y\dfrac{1}{\nu}\dfrac{du_y(x)}{d x}.
$$\smallskip

The dimension stream of momentum $p_{xy}$ equals
$$
p_{xy}=\dfrac{mn}{\beta}P_{xy}=
\dfrac{mn}{\beta}\dfrac{1}{\pi^{3/2}}\int \exp(-C^2)
C_xC_yh(x,{\bf C})d^3C,
$$
or
$$
p_{xy}=-\dfrac{kTn}{\nu}\dfrac{1}{2-\omega}\dfrac{du_y(x)}{dx}=
-\eta\dfrac{du_y(x)}{dx}.
$$

From here we find dynamic viscosity of gas
$$
\eta=\dfrac{kTn}{\nu}\dfrac{2}{2-\omega}.
\eqno{(2.2)}
$$
Kinematic viscosity is equal
$$
\nu_*=\dfrac{\eta}{\rho}=\dfrac{kT}{m\nu}\dfrac{2}{2-\omega}=
\dfrac{2D}{2-\omega}.
\eqno{(2.2')}
$$

Thus, the parameter $ \omega $ is defined by the equation (2.2) or
$(2.2')$:
$$
\omega=2\Big(1-\dfrac{\rho}{\eta}D\Big)=2\Big(1-\dfrac{D}{\nu_*}\Big).
\eqno{(2.3)}
$$

Now we will find  parameter $ \gamma $.
Let's consider now heat conduc\-tivity process.
Let in gas there is a constant gradient of temperature
$$
K_{T}=\dfrac{dT}{dz}=\const.
$$

Then the equation (1.7) transforms to the form
$$
C_z\dfrac{d\ln T}{dz}\Big(C^2-\dfrac{5}{2}\Big)=
\nu\sqrt{\beta}\Big[\gamma\Big(C^2-\dfrac{5}{2}\Big)Q_zC_z
-h(x,{\bf C})\Big],
\eqno{(2.4)}
$$
where
$\mathbf{Q}(\mathbf{r})$ is the dimensionless vector of heat
stream,
$$
\mathbf{Q}(\mathbf{r})=\dfrac{1}{\pi^{3/2}}\int \exp(-C^2)\mathbf{C}
\Big(C^2-\dfrac{5}{2}\Big)h(\mathbf{r},\mathbf{C})d^3C.
$$

Multiplying the equation (2.4) on
$$
\dfrac{1}{\pi^{3/2}}C_z\Big(C^2-\dfrac{5}{2}\Big)\exp(-C^2)
$$
and integrating, we receive
$$
\dfrac{5}{4}\dfrac{d\ln
T}{dz}=\nu\sqrt{\beta}\big(\dfrac{5}{4}\gamma-1\big)Q_z.
$$

The vector of a heat stream is connected with a temperature gradient
by equality
$$
q_z=-\varkappa \dfrac{dT}{dz},
$$
where $\varkappa$ is the coefficient of heat conductivity of gas,

$$
{q}_x(x)=\dfrac{n_0}{\pi^{3/2}}\int \exp(-C^2){v}_x
\Big(\dfrac{mv^2}{2}-\dfrac{5kT_0}{2}\Big)h(x,\mathbf{C})d^3C=$$$$=
\dfrac{n_0kT_0}{\sqrt{\beta}}{Q}_x(x).
$$

From last equalities we receive
$$
\dfrac{5}{4}\dfrac{d\ln
T}{dz}=\nu\sqrt{\beta}\big(\dfrac{5}{4}\gamma-1\big)Q_z=
\dfrac{\nu\beta}{nkT}\big(\dfrac{5}{4}\gamma-1\big)q_z
=$$$$=-\dfrac{\nu\beta}{nkT}\big(\dfrac{5}{4}\gamma-1\big)
\varkappa \dfrac{dT}{dz}.
$$

Whence we find
$$
\varkappa
=-\dfrac{5}{4}\dfrac{nk}{\nu\beta\big({5}\gamma/{4}-1\big)}.
\eqno{(2.5)}
$$

On the basis of (2.5) Prandtl's number for the equation (1.7) is
equal
$$
{\rm Pr}=\dfrac{5}{4\beta T}\dfrac{\eta}{\varkappa}=-\dfrac{5}{4\beta T}
\dfrac{kTn}{\nu}\dfrac{2}{2-\omega}\dfrac{4}{5}
\dfrac{\nu\beta\big({5}\gamma/{4}-1\big)}{nk}=$$$$=
\dfrac{-{5}\gamma/{2}+2}{2-\omega}.
$$

From last equation we find parameter $\gamma$
$$
\gamma=\dfrac{4}{5}\Big[1-\Big(1-\dfrac{\omega}{2}\Big)\Pr\Big],
\eqno{(2.6)}
$$
or
$$
\gamma=\dfrac{4}{5}\Big(1-\dfrac{D}{\nu_*}\Pr\Big).
$$

So, equation parameters $ \nu, \gamma, \omega $ are expressed
by equations (2.1), (2.3) and (2.6).

If we enter Ferziger's number
$
{\rm Fe}=\dfrac{D}{\nu_*}
$,
then relations (2.6) and (2.3) can be copied in the form
$$
\gamma=\dfrac{4}{5}\Big(1-{\Pr}{\rm Fe}\Big),
$$
$$
\omega=2\Big(1-{\rm Fe}\Big).
$$

Dependence of parameter $ \gamma $ from $ \omega $ represents
the straight line laying in second octant (fig. 1 see). At change
parameter $ \omega $ on range $ [2 (1-1/\Pr), 0] $ parameter $ \gamma $
runs the range $ [0,0.8 (1-\Pr)] $.

\begin{figure}
\begin{center}
\includegraphics[width=17.0cm, height=12cm]{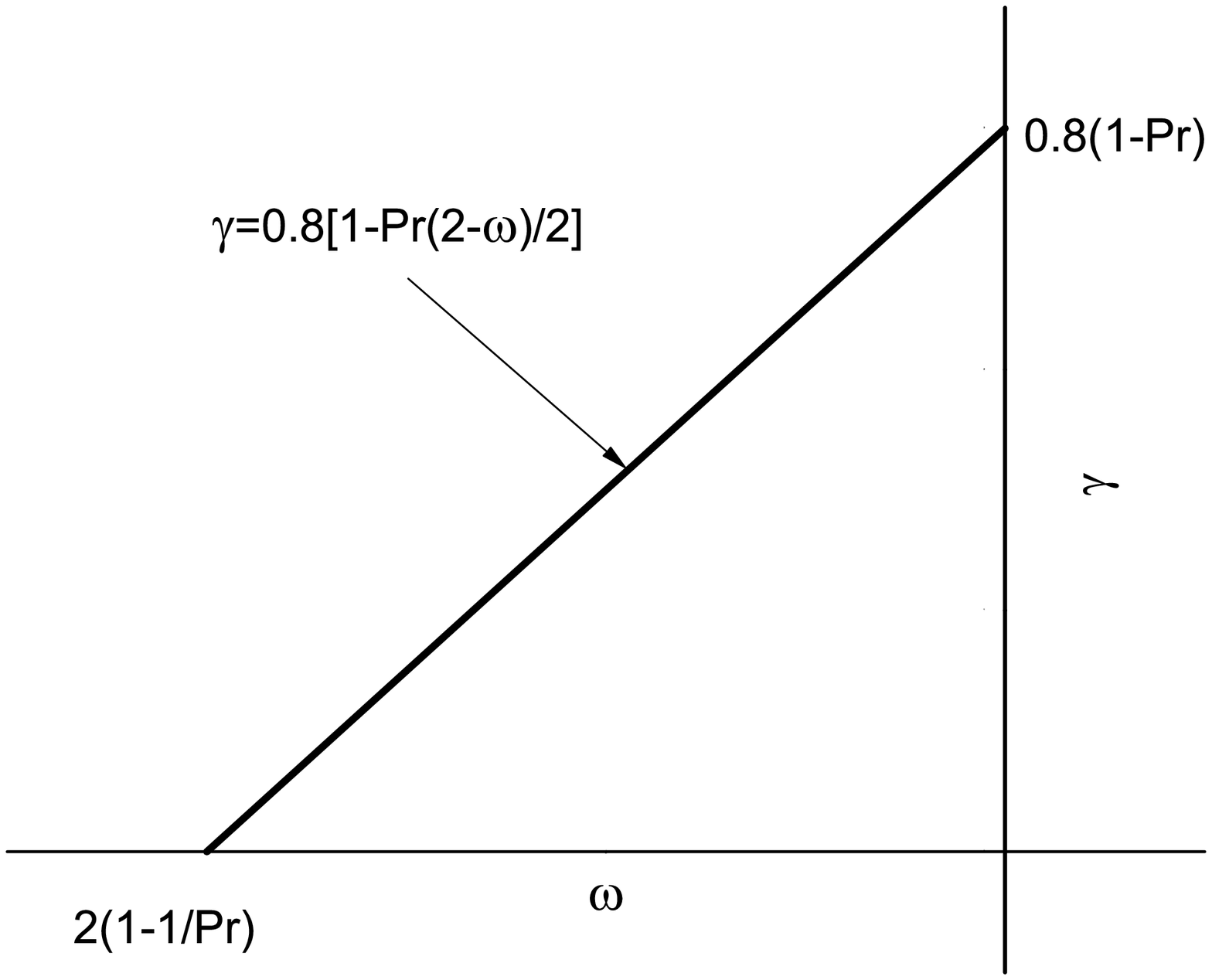}
Fig. 1.  Segment on straight line $ \gamma=0.8 [1-\Pr (1-\omega/2)] $,
on which Prandtl's number has constant value.
\end{center}
\end{figure}

\newpage

\begin{center}
\bf  4. Expression of parameters of the equation
through integral brackets
\end{center}

Parametres of the kinetic equation we will express through the integral
brackets \cite{Fer} in the first and the second approximations. As the second
approximation we will use approach of Kihara.

The self-diffusion coefficient is expressed through gas parametres in the first
approximation as follows \cite{Fer}.

$$
D^{(1)}=\dfrac{3}{8nm}\dfrac{\sqrt{\pi mkT}}{\pi \sigma^2}\cdot
\dfrac{1}{\Omega^{(1,1)*}}.
\eqno{(4.1)}
$$

In the second approximation for self-diffusion coefficient we have

$$
D^{(2)}=D^{(1)}\Big[1+\dfrac{(6C^*-5)^2}{40+32A^*}\Big].
\eqno{(4.2)}
$$

In expressions (4.1) and (4.2) the following standard designations
are accepted  \cite{Fer}

$$
A^*=\dfrac{\Omega^{(2,2)*}}{\Omega^{(1,1)*}},\qquad
C^*=\dfrac{\Omega^{(1,2)*}}{\Omega^{(1,1)*}},
$$

Here

$$
\Omega^{(i,j)*}=\dfrac{\Omega^{(i,j)}}{\big[\Omega^{(i,j)}\big]_{r.s.}}.
$$

Here the denominator represents $\Omega $--integral for model
molecules -- rigid spheres.

Equalities (4.1) and (4.2) give the chance to express the effective
collisions frequency in the first and the second approximation
through $ \Omega $--integrals. According to (2.1) we nave

$$
\nu^{(1)}=\dfrac{kT}{m}\cdot\dfrac{1}{D^{(1)}}=
\dfrac{8}{3}\sqrt{\pi}\sigma^2 n \sqrt{\dfrac{kT}{m}}\Omega^{(1,1)*}
\eqno{(4.3)}
$$
and
$$
\nu^{(2)}=\dfrac{kT}{m}\cdot\dfrac{1}{D^{(2)}}=
\nu^{(1)}\Big[1+\dfrac{(6C^*-5)^2}{40+32A^*}\Big]^{-1}.
\eqno{(4.4)}
$$ \medskip

As the first approximation coefficient of self-diffusion and
dynamic viscosity of gas are connected by the relation

$$
\dfrac{\rho }{\eta^{(1)}}D^{(1)}=\dfrac{6}{5}A^*.
\eqno{(4.5)}
$$

On the basis of (4.5) it is found dynamic viscosity of gas in the first
approximation (see \cite{Fer})

$$
\eta^{(1)}=\dfrac{5}{16}\dfrac{\sqrt{\pi mkT}}{\pi \sigma^2}
\cdot\dfrac{1}{\Omega^{(2,2)*}}.
\eqno{(4.6)}
$$

The second approximation of dynamic viscosity is equal

$$
\eta^{(2)}=
\eta^{(1)}\Big[1+\dfrac{3}{49}
\Big(\dfrac{\Omega^{(2,3)}}{\Omega^{(2,2)}}-\dfrac{7}{2}\Big)^2\Big].
\eqno{(4.7)}
$$

On the basis of (4.6) and (4.7) it is found kinematic viscosity in
the first and the second approximation

$$
\nu_*^{(1)}=\dfrac{\eta^{(1)}}{\rho}=\dfrac{5D^{(1)}}{6A^*}=
\dfrac{5}{16}\sqrt{\dfrac{\pi kT}{mn^2}}\dfrac{1}{\pi \sigma^2}
\cdot\dfrac{1}{\Omega^{(2,2)*}}
\eqno{(4.8)}
$$
and
$$
\nu_*^{(2)}=\dfrac{\eta^{(2)}}{\rho}=\dfrac{\eta^{(1)}}{\rho}
\Big[1+\dfrac{3}{49}
\Big(\dfrac{\Omega^{(2,3)}}{\Omega^{(2,2)}}-\dfrac{7}{2}\Big)^2\Big]=$$$$=
\nu_*^{(1)}\Big[1+\dfrac{3}{49}
\Big(\dfrac{\Omega^{(2,3)}}{\Omega^{(2,2)}}-\dfrac{7}{2}\Big)^2\Big].
\eqno{(4.9)}
$$

Let's find the Ferziger's number. As the first approximation
Ferziger's number on the basis of (4.1) and (4.8) it is equal

$$
{\rm Fe}^{(1)}=\dfrac{D^{(1)}}{\nu_*^{(1)}}=\rho\dfrac{D^{(1)}}{\eta^{(1)}}=
\dfrac{6}{5}A^*=\dfrac{6\Omega^{(2,2)*}}{5\Omega^{(1,1)*}}.
\eqno{(4.10)}
$$

 \newcommand{\Fe}{\mathop{\rm Fe\,}}
The Ferziger's number  on the basis of (4.9)  in the second approach
is equal

$$
{\Fe}^{(2)}={\Fe}^{(1)}\cdot
\dfrac{1+\dfrac{(6C^*-5)^2}{40+32A^*}}
{1+\dfrac{3}{49}
\Big(\dfrac{\Omega^{(2,3)}}{\Omega^{(2,2)}}-\dfrac{7}{2}\Big)^2}.
$$

Let us consider the Prandtl number. For this purpose it is required
heat conductivity of gas in the first and the second
approximations. Heat conductivity as a first approximation looks like
(see \cite{Fer})

$$
\varkappa^{(1)}=\dfrac{25}{32}\dfrac{\sqrt{\pi mkT}}{\pi \sigma^2
}\cdot c_v\cdot \dfrac{1}{\Omega^{(2,2)*}}.
$$
and in second approximation
$$
\varkappa^{(2)}=\varkappa^{(1)}\Big[1+\dfrac{2}{21}
\Big(\dfrac{\Omega^{(2,3)}}{\Omega^{(2,2)}}-\dfrac{7}{2}\Big)^2\Big].
$$

For the first approach it is received
$$
{\Pr}^{(1)}=\dfrac{5}{4\beta T}\cdot\dfrac{\eta^{(1)}}{\varkappa^{(1)}}
=\dfrac{2}{3},
$$
and in second approximation
$$
{\Pr}^{(2)}=\dfrac{5}{4\beta T}\cdot\dfrac{\eta^{(2)}}{\varkappa^{(2)}}=
\dfrac{2}{3}\cdot \dfrac{1+\dfrac{3}{49}
\Big(\dfrac{\Omega^{(2,3)}}{\Omega^{(2,2)}}-\dfrac{7}{2}\Big)^2}
{1+\dfrac{2}{21}
\Big(\dfrac{\Omega^{(2,3)}}{\Omega^{(2,2)}}-\dfrac{7}{2}\Big)^2}.
$$\medskip

Let's remind, that for model molecules -- rigid spheres \cite{Fer}

$$
\big[\Omega^{(l,r)}\big]_{r.s.}=
\Big(\dfrac{kT}{\pi m}\Big)^{1/2}\dfrac{(r+1)!}{2}
\Big[1-\dfrac{1+(-1)^l}{2(l+1)}\Big]\pi \sigma^2.
$$

Let us write out expressions for kinetic coefficient in case of
the gas consisting from molecules - rigid spheres. We will notice,
that according to definition
$$
A^*=1, \qquad C^*=1.
$$

Self-diffusion coefficient in the first and second approach
it is equal accordingly

$$
D^{(1)}=\dfrac{3}{8nm}\dfrac{\sqrt{\pi mkT}}{\pi \sigma^2}, \qquad
D^{(2)}=D^{(1)}\cdot 1.01388(8).
$$

Hence, for effective collisions  frequency in the first and
the second approximations it is received
$$
\nu^{(1)}=\dfrac{8}{3}\sqrt{\pi}\sigma^2 n \sqrt{\dfrac{kT}{m}},\qquad
\nu^{(2)}=\nu^{(1)}\cdot 0.986301.
$$\medskip

Dynamic viscosity of gas in the first and the second
approximations  it is equal accordingly
$$
\eta^{(1)}=\dfrac{5}{16}\dfrac{\sqrt{\pi mkT}}{\pi \sigma^2},\qquad
\eta^{(2)}=\eta^{(1)}\cdot 1.015306.
$$

Kinematic viscosity of gas in the first and the second approximations
it is  equal accordingly
$$
\nu_*^{(1)}=\dfrac{\eta^{(1)}}{\rho}=\dfrac{5D^{(1)}}{6A^*}=
\dfrac{5}{16}\sqrt{\dfrac{\pi kT}{mn^2}}\dfrac{1}{\pi \sigma^2},
\qquad\nu_*^{(2)}=\nu_8^{(1)}\cdot 1.015306.
$$ \medskip

The Ferziger's number  in the first and the second approximations
accor\-ding\-ly equals

$$
{\rm Fe}^{(1)}=\dfrac{6}{5},\qquad {\rm Fe}^{(2)}={\rm Fe}^{(1)}\cdot
0.998604.
$$

Prandtl number in the first and the second approximations
is  equal accordingly

$$
{\Pr}^{(1)}=\dfrac{2}{3}=0.66666(6), \qquad {\Pr}^{(2)}=\dfrac{2}{3}\cdot
0.991694=0.661129.
$$

For comparison we will write the expression of Prandtl number
cal\-cu\-lated according by the Chapman---Enskog method:
$ \Pr =\dfrac {2}{3} \cdot 0.990109=0.660073$.
Distinction between Prandtl number, calculated according by Kihara method
and Chapman---Enskog method, makes 0.16 \%.

\begin{center}
  {\bf 3. Conclusion}
\end{center}

In the present work the linear  kinetic Holway---Shakhov
equation for the rarefied one-atomic gas
is constructed.

This equation contains the members describing the relative
change of numerical density of gas, mass velocity of gas,
relative change of temperature of gas, heat stream in gas and
viscous stress tensor.

Numerical parameters of this equation are expressed through
coeffi\-ci\-ent of self-diffusion, kinematic (or dynamic) viscosity of gas
and heat conduc\-tivity.
Then these quantities are expressed through integral brackets.

Further authors purpose to present the analytical solution of
classical problems of the kinetic theory such as the
Kramers' problems about
isothermal sliding, the Maxwell problems about heat
sliding, the  Smo\-luc\-hov\-sky  problems about temperature jump and weak
evaporation.

\end{document}